\documentclass[twocolumn,pra,showpacs,amsmath,amssymb]{revtex4}

\usepackage{graphicx}

\def\be{\begin{equation}}
\def\ee{\end{equation}}
\def\bea{\begin{eqnarray}}
\def\eea{\end{eqnarray}}
\def\bma{\begin{mathletters}}
\def\ema{\end{mathletters}}

\def\0{\overline{0}}

\def\q0{\underline{0}}

\def\tr{\mbox{tr}}
\def\one{\leavevmode\hbox{\small1\normalsize\kern-.33em1}}

\def\bra#1{\langle#1|} \def\ket#1{|#1\rangle}

\def\proj#1{\ket{#1}\!\bra{#1}}

\begin{document}

\title{Non-secret correlations can be used to distribute secrecy}

\author{Joonwoo Bae$^{1}$, Toby Cubitt$^{2}$ and Antonio Ac\'{\i}n$^{3,4}$}
\affiliation{$^1$School of Computational Sciences, Korea Institute for
Advanced Study, Seoul 130-012, Korea\\
$^2$Department of Mathematics, University of Bristol, University Walk,
BS8 1TW, UK\\
$^3$ICFO-Institut de Ci\`encies Fot\`oniques,
Mediterranean Technology Park, 08860 Castelldefels (Barcelona), Spain\\
$^4$ICREA-Instituci\'o Catalana de Recerca i Estudis
Avan\c cats, Lluis Companys 23, 08010 Barcelona, Spain}
\date{\today}


\begin{abstract}
  A counter-intuitive result in entanglement theory was shown in [PRL
  \textbf{91} 037902 (2003)], namely that entanglement can be
  distributed by sending a separable state through a quantum
  channel. In this work, following an analogy between the entanglement
  and secret key distillation scenarios, we derive its classical
  analog: secrecy can be distributed by sending non-secret
  correlations through a private channel. This strengthens the close
  relation between entanglement and secrecy.
\end{abstract}

\maketitle

\section{Introduction}
Entangled and secret bits are different information resources that
turn out to be closely connected. An entangled bit, or \emph{ebit},
corresponds to a maximally entangled state of two qubits,
\begin{equation}\label{ebit}
    \ket{\Phi^+}=\frac{1}{\sqrt 2}(\ket{00}+\ket{11}) ,
\end{equation}
and represents the basic unit of bipartite entanglement \cite{BBPS}. A
standard problem in entanglement theory is, given an arbitrary
bipartite quantum state, $\rho_{AB}$, to determine how many ebits are
required for its formation or can be distilled out of it by local
operations and classical communication (LOCC).

On the other hand, secret bits, or \emph{sbits}, are the basic unit of
classical secret correlations. These arise when two honest parties, Alice
and Bob, share correlated random variables, $A$ and $B$, whereas the
eavesdropper, Eve, holds a third random variable $E$. The total
correlations are then described by a tripartite probability distribution,
$P(A,B,E)$. This distribution is a perfect sbit whenever
\begin{align*}
  P(A=B=0) &= P(A=B=1)=\frac{1}{2}\\
  P(A,B,E) &= P(A,B)P(E) .
\end{align*}
Note that Alice and Bob's variables are perfectly correlated, while
Eve gets no information whatsoever about them from her
outcome. Similarly to the case of quantum states, a basic question is
to quantify the number of sbits that are required to create a given
tripartite probability distribution $P(A,B,E)$, or that can be
distilled out of it by local operations and \emph{public}
communication (LOPC).

There exist several analogies between the entanglement properties of
quantum states and the cryptographic properties of the classical
probability distributions derived from them by local measurements
\cite{CP,GW}. In order to construct these analogies, one has to
explicitly introduce a third party in the quantum scenario. This can
easily be done by noting that any bipartite mixed state $\rho_{AB}$
can be seen as a tripartite pure state $\ket{\psi}_{ABE}$, such that
$\tr_E\proj{\psi}_{ABE}=\rho_{AB}$.  Indeed, the environment (that is,
the part of the global system that is not under the honest parties'
control) can naturally be associated with an adversary party, the
eavesdropper. The goal is then to connect the entanglement properties
of $\rho_{AB}$ to the cryptographic properties of those probability
distributions $P(A,B,E)$ that can be written as
\begin{equation}\label{meas}
    P(A,B,E)=\tr(M_A\otimes M_B\otimes M_E\proj{\psi}_{ABE}).
\end{equation}
$M_A$, $M_B$ and $M_E$ are positive operators defining a quantum
measurement in each local space, i.e. $\sum_i M_i=\one_i$ with
$i=A,B,E$. Of course, the same rule can be applied in a multipartite
scenario, where the quantum state $\rho_{ABC\ldots}$ is shared among $N$
parties.

A first rather trivial analogy follows from the fact that one sbit can
directly be obtained by measuring one ebit in, say, the computational
basis. This simple observation is behind some of the security proofs
of quantum key distribution protocols \cite{SP00}. Beyond this basic
analogy, other classical analogs of quantum information phenomena have
been derived, and vice-versa. For instance, the results of \cite{HOW}
on the existence of what is called negative quantum information were
translated into the classical scenario, obtaining analogous results
for the secret-key rate \cite{OSA}. In \cite{AG}, a systematic way of
mapping any entangled state onto a probability distribution containing
secret correlations was derived. One of the nicest concepts in this
direction is the existence of a cryptographic analog of bound
entanglement, known as bound information, first conjectured in
\cite{GW}. Recall that a quantum state is bound entangled when,
despite being entangled, it is impossible to distill pure ebits out of
it by LOCC. The existence and activation of non-distillable secret
correlations has been demonstrated in \cite{ACM} for the multipartite
scenario, adapting some known results for multipartite bound entangled
states. The existence of bipartite bound information remains an open
question. Other results have followed the opposite direction, going
from classical to quantum: the so-called squashed entanglement is an
entanglement measure whose construction was inspired by a known upper
bound on the secret-key rate \cite{CW}. In general, the connection
between entangled states and secret correlations is a useful tool in
the study of ebits and sbits, since it provides much insight into
these two fundamental resources. Note, however, that this analogy is
not a strict correspondence: there are ``exceptions'', such as the
existence of bound entangled states that can be mapped into
distillable probability distributions \cite{HHHO}.

A remarkable result in entanglement theory was obtained in
\cite{CVDC}, where it was shown that entanglement can be distributed
by sending a separable, non-entangled state through a quantum
channel. The scope of this work is to study whether a similar result
holds for probability distributions. Following the analogy between
ebits and sbits, we indeed prove that this is the case: secret
correlations can be distributed by sending non-secret correlations
through a private channel.

The article is structured as follows. In the next section, we
introduce the basic rules that apply when translating results from the
quantum to the classical scenario, and
vice-versa. Section~\ref{qreview} briefly reviews the results of
\cite{CVDC}, showing how to entangle two distant parties by sending a
separable state. The main results are given in section~\ref{mainsec},
where we show how to distribute secrecy by sending non-secret
correlations. Finally, we discuss some relevant issues and conclude.

\section{Entanglement vs. Secret Correlations}
\label{trules}
A standard scenario in entanglement theory consists of $N$ distant
parties, $A,B,C\ldots$, who share quantum correlations described by a
state $\rho$. The state may be mixed due to coupling to the
environment, $E$, the overall state being $\ket{\Psi}$, where
$\rho=\tr_E\proj{\Psi}$. The two main questions in this scenario are:
(i) is the preparation of $\rho$ possible by LOCC? and (ii) if not,
can pure ebits be distilled from $\rho$ by LOCC? These two questions
define the separability and distillability problems. When considering
the key-agreement scenario, many similarities appear (see for instance
\cite{CP,GW}). Now, $N$ distant honest parties and an eavesdropper
share correlated random variables, described by a probability
distribution $P(A,B,C,\ldots,E)$. The corresponding questions are: (i)
can these correlations be established by LOPC? and (ii) if not, can
pure sbits be distilled by LOPC?

Most of the analogies between the two scenarios can be summarised
as follows:
\begin{flushleft}
\begin{tabular}{ccc}
  quantum entanglement & \vline & secret correlations \\
  quantum communication & \vline & private communication \\
  classical communication & \vline & public communication \\
  local operations & \vline & local actions \\
\end{tabular}
\end{flushleft}
Here a private channel is a classical channel that is only accessible
to the honest parties. Using these intuitive rules, one can often
adapt results from entanglement theory to the secret correlations
scenario and vice-versa. For instance, a state is bound entangled
whenever its formation by LOCC is impossible but nevertheless it
cannot be transformed into pure ebits by LOCC. The corresponding
concept for secret correlations, knows as bound information, is simply
given by a probability distribution that, despite its formation by
LOPC being impossible, cannot be transformed into pure sbits by LOPC.

More quantitative statements can be made in the bipartite case. In the
case of quantum states, the ebit represents the basic unit of
entanglement. The number of ebits per copy that can be distilled out
of copies of a given quantum state by LOCC defines the distillable
entanglement, $E_{D}$ \cite{BDSW}. The corresponding classical analog
is the secret-key rate $S(A:B\|E)$ which gives the number of sbits
distillable from $P(A,B,E)$ by LOPC \cite{UW}. In a similar way, the
number of ebits required per copy for the formation of an entangled
state defines the entanglement cost, $E_C$. The so-called information
of formation, $I_{c}(A;B|E)$, introduced in \cite{RW}, represents its
classical analog. A probability distribution contains secret
correlations if, and only if, its information of formation is non-zero
\cite{RW}.

A useful upper bound for $S(A:B\|E)$ is given by the so-called
intrinsic information \cite{UW}. The intrinsic information between $A$
and $B$ given $E$ is defined as:
\begin{equation}\label{intrinf}
  I(A:B\downarrow E) = \min_{E\rightarrow \tilde{E}} I(A:B|\tilde{E}),
\end{equation}
where the minimisation runs over all possible stochastic maps
$P(\tilde{E}|E)$ defining a new random variable $\tilde{E}$. The
quantity $I(A:B|E)$ is the mutual information between $A$ and $B$
conditioned on $E$. It can be written as
\begin{equation*}
  I(A:B|E) = H(A,E)+H(B,E)-H(A,B,E)-H(E),
\end{equation*}
where $H(X)$ is the Shannon entropy of the random variable $X$. It
also gives a lower bound on the information of formation \cite{RW},
thus
\begin{equation}\label{ibounds}
  S(A:B\parallel E)\leq I(A:B\downarrow E)\leq I_{c}(A;B|E).
\end{equation}
In fact, $I_{c}(A;B|E)>0$ if, and only if, $I(A:B\downarrow E)>0$
\cite{RW}. The intrinsic information plays a key role in the proof of
our results.

\section{Quantum Scenario}
\label{qreview}
Before presenting our results, we summarise the findings of
Ref.~\cite{CVDC} on the distribution of entanglement by means of a
separable state. The scenario consists of two initially uncorrelated
distant parties who are connected by a classical and a quantum channel
\cite{note}. In order to entangle two distant qubits, $A$ and $B$, the
parties must use the quantum channel, since no entanglement can be
created by LOCC. Thus one of the parties, say Alice, should prepare an
additional qubit, $C$, and send it to Bob. Clearly, a sufficient
condition for entanglement distribution is that the mediating quantum
particle $C$ is entangled with Alice's quantum system $A$, so that Bob
becomes entangled with her after receiving it. Intuitively, one would
expect that this is also a necessary condition. Remarkably, this is
not the case, as shown in Ref.~\cite{CVDC}, where an explicit
counterexample is provided in which Alice distributes entanglement to
Bob by sending a qubit $C$ through the quantum channel that is never
entangled across the partition $C-AB$.

The example works as follows. Alice holds two qubits, $A$ and $C$,
while Bob has one qubit, $B$, in the initial state
\begin{equation}\label{initials}
  \rho_{ABC} = \frac{1}{6}\sum_{k=0}^3 \proj{\Psi_k,\Psi_{-k},0}
              + \sum_{i=0}^1 \frac{1}{6}\proj{i,i,1}
\end{equation}
where $\ket{\Psi_{k}} = (\ket{0} + e^{i\pi k/2}\ket{1})
/\sqrt{2}$. This state is fully separable across all partitions, so it
can be prepared by LOCC. Alice now applies a controlled-NOT (CNOT)
operation to her qubits, where $A$ ($C$) is the control (target)
qubit, resulting in the state
\begin{equation}
  \sigma_{ABC} = \frac{1}{3} \proj{\Psi_{GHZ}}
                + \sum_{i,j,k=0}^1 \beta_{ijk}\proj{ijk}
  \label{eq:GHZ}
\end{equation}
where $\ket{\Psi_{GHZ}}_{ABC} =
\left(\ket{000}+\ket{111}\right)/\sqrt{2}$, $\beta_{001} = \beta_{010}
= \beta_{101} = \beta_{110} = 1/6$, and all other $\beta_{ijk} = 0$.
This state is still separable across the $C-AB$ partition \cite{DC}.
Alice now sends $C$ to Bob, who applies a CNOT with $B$ ($C$) as the
control (target) qubit. After all these steps, Alice and Bob share a
state
\begin{equation}
    \tau_{ABC}
    = \frac{1}{3}\proj{\Phi^+}_{AB}\otimes\proj{0}_C
      + \frac{2}{3}\one_{AB}\otimes\proj{1}_C ,
\end{equation}
where Bob has both $B$ and $C$. This state is distillable. Indeed, by
measuring particle $C$ in the computational basis, Alice and Bob's
systems are projected into a maximally entangled state of two qubits
with probability $1/3$.

\section{A Translated Classical Scenario}
\label{mainsec}
We now translate the previous quantum result to the key-agreement
scenario. Namely, we show that secret correlations can be distributed
by sending through the private channel a random variable that does not
have secret correlations with either Alice and/or Bob. For the
construction of the example, we can follow the ``rules'' given in
section~\ref{trules}. Following~\eqref{meas}, the initial quantum
state~\eqref{initials} is replaced by the probability distribution
obtained by measuring in the computational bases:
\begin{equation}\label{initialp}
  \begin{array}{ccccc}
    \hline \hline
    A & B & C & E & P(A,B,C,E) \\
    \hline
    0 & 0 & 0 & e_{0} & 1/6 \\
    0 & 1 & 0 & e_{01} & 1/6 \\
    1 & 0 & 0 & e_{10} & 1/6 \\
    1 & 1 & 0 & e_{0} & 1/6 \\
    0 & 0 & 1 & f_{0} & 1/6 \\
    1 & 1 & 1 & f_{1} & 1/6 \\
    \hline \hline
  \end{array}
\end{equation}
Recall that for any separable state there exists an optimal
measurement by Eve such that the intrinsic information for the
obtained distribution~\eqref{meas} is zero for all choices of measurements by
Alice and Bob \cite{GW}. However, Eve's optimal measurement is not
necessarily in the computational basis. Thus, it is not immediate that the distribution~\eqref{initialp} contains no secret
correlations. However, it is possible to prove that this is indeed the case: the distribution has zero intrinsic information
across the bipartition $AC-B$, since $I(AC:B|E)=0$, which \emph{does}
imply that Alice and Bob do not share secret correlations.

Now, Alice performs the (classical) CNOT operation on $A$ and $C$, and
sends $C$ through the private channel to Bob, who performs the CNOT
operation on $B$ and $C$. After Alice's CNOT, the probability
distribution is
\begin{equation}\label{mid}
  \begin{array}{cccccc}
    \hline \hline
    A & B & C & E & P(A,B,C,E) \\
    \hline
    0 & 0 & 0 & e_{0} & 1/6 \\
    0 & 1 & 0 & e_{01} & 1/6 \\
    1 & 0 & 1 & e_{10} & 1/6 \\
    1 & 1 & 1 & e_{0} & 1/6 \\
    0 & 0 & 1 & f_{0} & 1/6 \\
    1 & 1 & 0 & f_{1} & 1/6 \\
    \hline \hline
  \end{array}
\end{equation}
This distribution has zero intrinsic information across the partition
$C-AB$. Indeed, consider the map $E\to\bar{E}$ in which Eve replaces
$f_{0}$ and $f_{1}$ by $e_{0}$, but leaves everything else
untouched. The resulting probability distribution has $I(A:B|E)=0$,
thus $I(AB:C\downarrow E) = 0$ for~\eqref{mid}. That is, the $C$ that
is sent through the private channel does not share secret correlations
with $A$ and/or $B$.

The final probability distribution between Alice and Bob, after Bob's
CNOT, is
\begin{equation}\label{final}
  \begin{array}{ccccc}
    \hline \hline
    A & B & C & E & P(A,B,C,E) \\
    \hline
    0 & 0 & 0 & e_{0} & 1/6 \\
    0 & 1 & 1 & e_{01} & 1/6 \\
    1 & 0 & 1 & e_{10} & 1/6 \\
    1 & 1 & 0 & e_{0} & 1/6 \\
    0 & 0 & 1 & f_{0} & 1/6 \\
    1 & 1 & 1 & f_{1} & 1/6 \\
    \hline \hline
  \end{array}
\end{equation}
We now show how Alice and Bob can distill 1 sbit from this
distribution. Bob holds two of the random variables, $B$ and $C$. He
receives $C=0$ and $C=1$ with probabilities, $1/3$ and $2/3$
respectively. By LOPC, Alice and Bob keep their outcome whenever
$C=1$, otherwise they reject the instance. Thus, with probability
$1/3$, Alice and Bob (and Eve) are correlated according to:
\begin{equation*}
  \begin{array}{ccccc}
    \hline \hline
    A & B & C & E &  P(A,B,C,E) \\
    \hline
    0 & 0 & 0 & e_{0} & 1/2 \\
    1 & 1 & 0 & e_{0} & 1/2 \\
    \hline \hline
  \end{array}
\end{equation*}
which defines a perfect sbit. Thus, Alice and Bob are able to
distribute distillable secret correlations by sending a random
variable, that does not itself have secret correlations, through a
private channel.

\section{Concluding Remarks}
\label{concl}
In this work, we have constructed the cryptographic analog of the
distribution of (distillable) entanglement by a separable state:
(distillable) secrecy can be distributed by sending non-secret
correlations. This result is completely equivalent to its entanglement
analog: the use of the quantum (private) channel is essential for the
successful entanglement (secrecy) distribution, even though the
mediating particle (random variable) never has quantum (secret)
correlations with the sender and/or receiver.

At first sight, the existence of this cryptographic analog suggests
some interesting possibilities. For instance, one might imagine that
secrecy could be distributed by an untrusted messenger, Charlie, who
after transmitting the relevant information could not break the
established secret key. Clearly, this is not the case if the
transmitter can later collaborate with the eavesdropper. Indeed, since
the information is classical, Charlie can keep a perfect copy of the
transmitted random variable, $C$, and give it to Eve. The channel is
no longer private, and it is known that the distribution of secret
correlations by LOPC is impossible.

One could however consider a less restrictive scenario, in which the
transmitter is still untrusted, but it is assumed that he does not
collaborate with Eve. Can Alice and Bob use the above protocol to
distill a secret key against Eve and Charlie separately? Indeed, they
can. However, a much simpler protocol already achieves this. Alice,
Bob and Eve initially share a public perfectly correlated bit,
$P(a=b=e=0)= P(a=b=e=1)=1/2$. Of course, no key extraction is
possible. Then, Alice generates a random bit which she sends to Bob,
via the messenger Charlie. Charlie leaves and may try to break the
key, but he is not allowed to collaborate with Eve. When he delivers
the random bit, Alice, Bob and Charlie also share a perfectly
correlated bit. It is clear that Alice and Bob can distill a key
against Charlie and Eve (if they don't collaborate), by taking the XOR
of their two bits.

\section{Acknowledgements}

This work is supported by the EU QAP project, the Spanish MEC, under
FIS2007-60182 and Consolider-Ingenio QOIT projects, the Caixa Manresa, the Generalitat de Catalunya, and the IT R\&D program of MIC/IITA [2005-Y-001-04 , Development of next generation security technology].


\begin{thebibliography}{18}
\bibitem{note}
This artificial distinction between the channels is made for the sake of clarity. Indeed, classical information can be sent through the quantum channel.
\end{thebibliography}
\end{document}